\documentclass[prl,aps,twocolumn,showpacs,preprintnumbers,amsmath,amssymb,
superscriptaddress,notitlepage]{revtex4-1}

\usepackage{graphicx}
\usepackage{dcolumn}
\usepackage{bm}
\usepackage{hyperref}

\font\msytw=msbm9 scaled\magstep1 \font\msytww=msbm7
scaled\magstep1

\let\a=\alpha \let\b=\beta    \let\d=\delta 
  \let\h=\eta   \let\th=\theta  \let\l=\lambda
\let\m=\mu    \let\n=\nu    \let\x=\xi     \let\p=\pi    \let\r=\rho
\let\s=\sigma     
   
\let\G=\Gamma   \let\L=\Lambda \let\X=\Xi
         
\let\O=\Omega 

 \def\VV{{\cal V}}
 
 \def\BBB{{\cal B}}
  
\def\DD{{\cal D}} \def\SS{{\cal S}}

   \def\pp{{\bf p}}
 \def\xx{{\bf x}} \def\yy{{\bf y}} 

\def\kk{{\bf k}}

\def\RRR{\hbox{\msytw R}} \def\rrrr{\hbox{\msytww R}}

 \def\ZZZ{\hbox{\msytw Z}}

\def\\{\hfill\break}
\let\==\equiv
\let\io=\infty
\let\0=\noindent
\def\media#1{{\langle#1\rangle}}
\let\dpr=\partial

\def\const{{\rm const}}
\def\tende#1{\,\vtop{\ialign{##\crcr\rightarrowfill\crcr\noalign{\kern-1pt
    \nointerlineskip} \hskip3.pt${\scriptstyle #1}$\hskip3.pt\crcr}}\,}
\def\otto{\,{\kern-1.truept\leftarrow\kern-5.truept\to\kern-1.truept}\,}

\def\to{\rightarrow}

\def\qed{\hfill\raise1pt\hbox{\vrule height5pt width5pt depth0pt}}

\def\Tr{{\rm Tr}}

\def\V#1{{\bf#1}}
\def\be{\begin{equation}}
\def\ee{\end{equation}}
\def\bea{\begin{eqnarray}}
\def\eea{\end{eqnarray}}
\def\nn{\nonumber}
\def\pref#1{(\ref{#1})}

\begin{document}

\title{Absence of interaction corrections in
graphene conductivity}
\author{Alessandro Giuliani}
\affiliation{%
Universit\`a di Roma Tre, L.go S. L. Murialdo 1, 00146 Roma,
Italy}
\author{Vieri Mastropietro}%
\affiliation{%
Universit\`a di Roma Tor Vergata, Viale della Ricerca Scientifica
00133 Roma, Italy}
\author{Marcello Porta}
\affiliation{%
Universit\`a di Roma La Sapienza, P.le A. Moro 2, 00185 Roma,
Italy}

\begin{abstract} The exact vanishing of the interaction corrections to the
zero temperature and zero frequency conductivity of graphene in
the presence of weak short range interactions is rigorously
established.
\end{abstract}
 \maketitle

Graphene \cite{N} has several peculiar properties originating from
its perfect two dimensionality and from the Dirac-like nature of
its charge carriers at half-filling. In particular, recent optical
measurements \cite{N1} show that at half-filling and small temperatures, if the
frequency is in a range between the temperature and the
band-width, the conductivity is essentially constant and equal, up
to a few percent, to $\s_0=\frac{e^2}{h}\frac{\pi}{2}$; such value only 
depends on the fundamental von Klitzing constant $h/e^2$ and
not on the material parameters, like the Fermi velocity. These
experimental results confirm the theoretical predictions \cite{A} based on
the description of graphene in terms of massless non-interacting Dirac
particles \cite{S,L}; lattice effects have been taken into account
in \cite{SPG}. Since truly universal phenomena are quite rare in
condensed matter (an example is provided by the Quantum Hall
effect), it is important to understand whether this apparently {\it
universal value} is just an artifact of the idealized description
in terms of non-interacting fermions or rather it is a robust
property still valid in the presence of electron-electron
interactions, which are certainly present and expected to play a
role in real graphene. This question is entirely analogous to the
one concerning universality in the quantum Hall effect \cite{IM},
a notoriously difficult and still open problem.

The effects of the electron-electron interactions on the graphene
conductivity have been investigated in the Dirac approximation by
perturbation theory both in the presence of long- and of
short-ranged interactions; however, lowest order explicit
computations have produced {\it different} results \cite{SS1,M1,
H0}, depending on the regularization scheme (momentum cut-off or
dimensional regularization) chosen to cure the spurious
ultraviolet divergences introduced by the Dirac approximation. In
\cite{M1} it was predicted that in the presence of electrostatic
interactions the zero frequency conductivity tends to zero, while in 
\cite{SS1,H0} it was argued that it converges to the free Dirac one,
as a consequence of the {\it divergence} of the Fermi
velocity; however, if screening or retardation effects are 
taken into account, the Fermi velocity is known to saturate at 
low frequency \cite{GGV,GMP,GM}, in which case it is unclear 
what to expect. The extreme sensitivity of the conductivity computation to
approximations or regularizations (see also \cite{Z}) calls for a
rigorous analysis.

In this paper we consider the Hubbard model on the honeycomb
lattice, as a model of monolayer graphene with screened
interactions. While in general the understanding of the low temperature behavior of 
the Hubbard model is a formidable challenge for theoreticians,
in the case of the honeycomb lattice at half filling the methods
introduced in \cite{GM} and based on constructive Renormalization
Group have proved to be quite effective. Using these techniques, we
rigorously establish {\it the exact (non-perturbative) vanishing
of the interaction corrections to the conducivity} in the zero
frequency limit. All Feynman graphs contributing to the
conductivity cancel out exactly in the limit, a statement
analogous to the Adler-Bardeen theorem in quantum electrodynamics
\cite{AB,IM}.

We introduce creation and annihilation fermionic operators
$\psi_{\vec x,\s}^\pm=(a^\pm_{\vec x,\s}, b^\pm_{\vec x + \vec
\d_1,\s})= L^{-2}\sum_{\vec k\in\BBB_\L}\psi^\pm_{\vec k,\s}
e^{\pm i\vec k\vec x}$ for electrons with spin index
$\s=\uparrow\downarrow$ sitting at the sites of the two triangular
sublattices $\L_A$ and $\L_B$ of a periodic honeycomb lattice of
side $L$; we assume that $\L_A=\L$ has basis vectors $\vec
l_{1,2}=\frac12(3,\pm\sqrt3)$ and that $\L_B=\L_A+\vec\d_j$, with
$\vec \d_1=(1,0)$ and $\vec\d_{2,3}=\frac12 (-1,\pm\sqrt3)$ the
nearest neighbor vectors; $\BBB_\L=\{\vec k=n_1\vec G_1/L+ n_2\vec
G_2/L\,:\, 0\le n_i<L\}$ with $\vec
G_{1,2}=\frac{2\p}3(1,\pm\sqrt3)$ is the first Brillouin zone
(note that in the thermodynamic limit $L^{-2}\sum_{\vec
k\in\BBB_\L}\to|\BBB|^{-1} \int_{\BBB} d\vec k$, with $|\BBB|
=8\pi^2/(3\sqrt{3})$). The grand-canonical Hamiltonian at
half-filling is $H_\L=H^0_\L+U V_\L$, where $H_0$ is the free
Hamiltonian, describing nearest neighbor hopping ($t$ is the
hopping parameter):
\be H^0_\L(t)= -t\sum_{\substack{\vec x\in\L_A \\
j=1,2,3}}\sum_{\s=\uparrow\downarrow}
(a^{+}_{\vec x,\s} b^{-}_{\vec x +\vec
\d_j,\s}+b^{+}_{\vec x +\vec
\d_j,\s}a^{-}_{\vec x,\s} )\label{1}\ee
and $V_\L$ is the local Hubbard interaction:
\be V_\L=\sum_{\vec x\in\L_A}\prod_{\s=\uparrow\downarrow}(a^+_{\vec x,\s}
a^-_{\vec x,\s}-\frac12)+\sum_{\vec x\in\L_B}\prod_{\s=\uparrow\downarrow}
(b^+_{\vec x,\s}b^-_{\vec x,\s}-\frac12)\;.\label{2}\ee
In order to define the lattice current and the conductivity, we
modify the hopping parameter along the bond $(\vec x,\vec
x+\vec\d_j)$ as $t\to t_{\vec x,j}(\vec A)=t\exp\{ie\int_0^1\vec
A(\vec x+s\vec\d_j)\cdot\vec\d_j\,ds\}$, where $\vec A(\vec x)\in\RRR^2$ is
a periodic continuum field on ${\cal S}_\L= \{\vec x=L\x_1\vec
l_1+L\x_2 \vec l_2\,:\, \x_i\in[0,1)\}$; its Fourier transform is
defined as $\vec A(\vec x)=|\SS_\L|^{-1}\sum_{\vec p\in\DD_\L}
\vec A_{\vec p}e^{-i\vec p\vec x}$, where
$|\SS_\L|=\frac{3\sqrt3}2L^2$ and $\DD_\L=\{\vec p=n_1\vec G_1/L+
n_2\vec G_2/L\,:\, n_i\in\ZZZ\}$ (note that in the thermodynamic
limit $|\SS_\L|^{-1}\sum_{ \vec
p\in\DD_\L}\to(2\p)^{-2}\int_{\rrrr^2} d\vec p$). If we denote by
$H(A)= H_\L^0(\{t_{\vec x,j}(\vec A)\})+UV_\L$ the modified
Hamiltonian with the new hopping parameters, the lattice current
is defined as $\vec \jmath_{\vec p}=-\frac{\dpr H(A)}{\dpr \vec A_{\vec
p}}$, which gives, at first order in $\vec A$, \be \vec
\jmath_{\vec p}=\vec \jmath_{\vec p}^{(P)}+\int \frac{d\vec
q}{(2\p)^2} \widehat \jmath^{(D)}_{\vec p,\vec q}\vec A_{\vec
q}\;,\label{1.3}\ee
where $\int \frac{d\vec q}{(2\p)^2}$ is a shorthand for $|\SS_\L|^{-1}\sum_{\vec q\in\DD_\L}$
and, if $\int_{\BBB} \frac{d\vec k}{|\BBB|}$ is a shorthand for $L^{-2}\sum_{\vec k\in\BBB_\L}$
and $\h^j_{\vec p}=\frac{1-e^{-i\vec p\vec \d_j}}{i\vec p\vec \d_j}$,
\bea \vec \jmath_{\vec p}^{(P)}&=& -iet\sum_{\s,j}
\int_{\BBB} \frac{d\vec k}{|\BBB|}
\big(a^+_{\vec k+\vec p,\s}b^-_{\vec k,\s}\vec \d_j \h^j_{\vec p}
e^{-i\vec k(\vec
\d_j-\vec\d_1)}\nn\\
&&-b^+_{\vec k+\vec p,\s}a^-_{\vec k,\s}\vec \d_j \h^j_{\vec p}
e^{+i(\vec k+\vec p)(\vec\d_j-\vec\d_1)}\big)\label{1.4}\eea
is the {\it paramagnetic current} and
\bea&& \big[\,\widehat \jmath_{\vec p,\vec q}^{(D)}\big]_{lm}= e^2 t\sum_{\s,j}
(\vec \d_j)_l
(\vec \d_j)_m \h^j_{\vec p}\h^j_{\vec q}\int_{\BBB} \frac{d\vec k}{|\BBB|}
\big(a^+_{\vec k+\vec p+\vec q,\s}b^-_{\vec k,\s}\nn\\&&
\cdot\, e^{-i\vec k(\vec
\d_j-\vec\d_1)}+b^+_{\vec k+\vec p+\vec q,\s}a^-_{\vec k,\s}
e^{i(\vec k+\vec p+\vec q)(\vec\d_j-\vec\d_1)}\big)\label{1.5}\eea
is the {\it diamagnetic tensor}.
The conductivity, at Matsubara frequency $p_0\in 2\p\b^{-1}(\ZZZ+\frac12)$
and in units such that $\hbar=1$, is defined via Kubo formula as
\cite{SPG}
\be \s^{\b,\L}_{lm}(p_0)=-\frac{K^{\b,\L}_{lm}(p_0,\vec 0)}{p_0 |\SS_\L|}\ee
where, if $\X=\Tr\{e^{-\b H_\L}\}$, $\media{\cdot}=\X^{-1}\Tr\{e^{-\b H_\L}\cdot\}$ and
$O_{x_0}=e^{H_\L x_0}O e^{-H_\L x_0}$ for a generic operator $O$,
\be K^{\b,\L}_{lm}(p_0,\vec p)=\int_0^\b \!\!\!dx_0\, e^{-i p_0 x_0}
\media{j^{(P)}_{x_0,\vec p,l} \,j^{(P)}_{0,-\vec p,m}}+
\media{\big[\,\widehat \jmath^{(D)}_{\vec p,-\vec p}\big]_{lm}}\label{1.6} \ee
It is known that in general the interaction {\it modifies} the values
of the physical quantities; for instance, the Fermi velocity $v_F$, the wave
function renormalization $Z$ and the vertex functions are known to depend explicitly
on the interaction \cite{GM}; moreover, it was proven in \cite{GM} that $v_F$, $Z$ and the vertex functions are {\it analytic} functions of $U$ for $|U|$ small enough, uniformly as $\b,|\L|\to\infty$. In this Letter we prove a similar result for the conductivity. Moreover, we prove that in the
thermodynamic, zero temperature and zero frequency limit, the conductivity
is {\it universal}, i.e., it is exactly independent of $U$.
\vskip.1cm {\bf Theorem.} {\it There exists a constant $U_0>0$ such that,
for $|U|\le U_0$ and any fixed $p_0$, $\s^{\b,\L}_{lm}(p_0)$ is analytic in $U$
uniformly in $\b,\L$ as $\b,|\L|\to\infty$. Moreover,
\be \lim_{p_0\to 0^+}\lim_{\b\to\io}\lim_{|\L|\to\infty}\s^{\b,\L}_{lm}(p_0)=
\frac{e^2}{h}\frac{\pi}{2}
\d_{lm}\;.
\label{1.7}\ee }
Note that the limit $\b\to\io$ is taken before the limit $p_0\to
0^+$. In other words, the theorem says that the interaction
corrections to the conductivity are negligible at frequencies
$\b^{-1}\ll p_0\ll t$.

\vskip.1cm {\bf Proof.} The idea of the proof is based on the two
main ingredients: (i) {\it exact lattice} Ward Identities (WI)
relating the current-current, vertex and 2-point functions; (ii) the
fact that the interaction-dependent corrections to the Fourier
transform of the current-current correlations are {\it
differentiable} with continuous derivative (in contrast, the free part is continuous
and not differentiable at zero frequency). This
last property follows from the non-perturbative estimates found in
\cite{GM}, which we now briefly recall. The generating
functional for correlations can be written in terms of a Grassmann
integral:
\be e^{W(A,\l)}=\int P(d\psi)
e^{\VV(\psi)+(\psi,\l)+B(A,\psi)}\label{por} \ee
where, if $\kk=(k_0,\vec k)$ with $k_0$ the Matsubara frequency,
$P(d\psi)$ is the fermionic gaussian integration for
$\psi^\pm_{\kk,\s}=(a^\pm_{\kk,\s},b^\pm_{\kk,\s})$,
with inverse propagator
\be g^{-1}(\kk)= -Z_0\left(\begin{array}{cc} ik_0 & v_0 \O^*(\vec k)
\\ v_0 \O(\vec k) & ik_0 \end{array}\right)\;,\label{1.8}\ee
with $Z_0=1$, $v_0=\frac32t$ and
$\O(\vec k) = \frac23\sum_{j=1,2,3}e^{i\vec k(\vec \d_j -
\vec\d_1)}$ (note that $g(\kk)$ is singular only at the Fermi
points $\kk=\kk_F^{\pm}
=(0,\frac{2\p}{3},\pm\frac{2\p}{3\sqrt3})$). Moreover, $B(A,\psi)=$
\bea&& =\sum_\s\int_0^\b dx_0\sum_{\vec x\in\L}\Big[-ie\psi^+_{\xx,\s}
\begin{pmatrix}A_0(\xx)&0\\0&A_0(\xx+\d_1)\end{pmatrix}\psi^-_{\xx,\s}
\nn\\
&&+\sum_j \big((t_{\vec x,j}(\vec A)-t)
a^+_{(x_0,\vec x),\s}b^-_{(x_0,\vec x+\vec\d_j),\s}+c.c.\big)\Big]
\label{1.9}\eea
and $(\psi,\l)=\int_0^\b dx_0\sum_{\vec x\in\L}
[\psi^+_\xx\l_\xx^- +\l^+_\xx\psi^-_\xx]$. The response function
$K^{\b,\L}(\pp)$ corresponds to the spatial components of the
tensor $\widehat K_{\m\n}(\pp)=\frac{\d^2}{\d A_\m(\pp) \d
A_\n(-\pp)} W(A,0)\big|_{A=0}$, with $\m,\n=0,1,2$. Performing the
phase transformation $\psi^{\pm}_\xx\to e^{\pm i e
\a_\xx}\psi^{\pm}_\xx$ in Eq.(\ref{por}), we find
\be W(A+\dpr\a,\l e^{ie\a})=W(A,\l)\;, \label{1.10}\ee
which implies the following lattice Ward Identity \cite{H}
\be \sum_{\m=0}^2 p_\m \widehat K_{\m\n}(\pp)=0\;,\label{1.11} \ee
for all $\n\in\{0,1,2\}$. On the other hand, the functional
integral Eq.(\ref{por}) can be evaluated in terms of an exact
Renormalization Group (RG) analysis, described in full detail in
\cite{GM}. We decompose the field $\psi$ as a sum of fields
$\psi^{(k)}$, living on momentum scales $|\kk-\kk_F^\pm|\simeq
2^{h}$, with $h\le 0$ a scale label; the iterative integration of
the fields on scales $h<h'\le 0$ leads to an effective theory
similar to Eq.(\ref{por}) with a cut-off around the
Fermi points of width $2^h$ and with a scale dependent propagator
$g^{(\le h)}(\kk)$ with the same singularity structure as
Eq.(\ref{1.8}), with $Z_0$ and $v_0$ replaced by $Z_h$ and $v_h$,
respectively (the effective wave function renormalization and
Fermi velocity on scale $h$). Moreover, setting for simplicity
$\l=0$, at scale $h$ the interaction $\VV(\psi)+B(A,\psi)$ is
replaced by an effective interaction $\VV^{(\le h)}(\psi^{(\le
h)})+B^{(h)}(A,\psi^{(\le h)})$, with the {\it effective
potential} $\VV^{(\le h)}(\psi^{(\le h)})$ a sum of monomials in
$\psi^{(\le h)}$ of arbitrary order, characterized at order $n$ by
kernels $W^{(h)}_{n,0}(\xx_1,\ldots,\xx_n)$ that are analytic in
$U$ and decay super-polynomially in the relative distances
$|\xx_i-\xx_j|$ on scale $2^{-h}$; moreover the {\it effective
source} is given by
\be B^{(h)}(A,\psi)=
\sum_{\m=0}^2 Z_{\m,h}\int \frac{d\pp}{(2\p)^3}A_{\m}(\pp)
j_{\m}(\pp)+\bar B^{(h)}\label{1.13} \ee
where $j_0(\pp)=-ie\sum_\s\int\frac{d\kk}{(2\p)|\BBB|}
\psi^+_{\kk+\pp,\s}\G_0(\vec k,\vec p)\psi^-_{\kk,\s}$,
$\vec \jmath(\pp)=-ie\sum_\s\int\frac{d\kk}{(2\p)|\BBB|}
\psi^+_{\kk+\pp,\s}\vec \G(\vec k,\vec p)\psi^-_{\kk,\s}$,
$[\G_0(\vec k,\vec p)]_{ij}=\d_{ij}\exp\{-ip_1\d_{i2}\}$ and
\be \vec\G(\vec k,\vec p)=\frac23\sum_j\vec \d_j \h^j_{\vec p}\begin{pmatrix}0 &
-e^{-i\vec k(\vec\d_j-\vec\d_1)}\\
e^{+i(\vec k+\vec p)(\vec\d_j-\vec\d_1)}&0\end{pmatrix}\;.\label{1.14}\ee
Finally, $\bar B^{(h)}$ is a sum of monomials in $(A,\psi)$ of
arbitrary order, characterized at order $n$ in $\psi$ and $m$ in
$A$ by kernels
$W^{(h)}_{n,m}(\xx_1,\ldots,\xx_n;\yy_1,\ldots,\yy_m)$ that are
analytic in $U$, decay super-polynomially in the relative
distances on scale $2^{-h}$ and are non-zero only if $m\ge 1, n\ge
0$ and $m+n\ge 3$; in particular, for all $0<\th<1$, they satisfy the
bounds (proved in \cite{GM}),
\bea && \int d\xx_2\cdots d\xx_n d\yy_1\ldots d\yy_m\,
|W^{(h)}_{n,m}(\xx_1,\ldots,\xx_n;\yy_1,\ldots,\yy_m)|\nn\\
&&\le (\const.) |e|^m2^{(3-n-m)h}\big(1-\d_{m,0}+|U|2^{\th
h}\big)\;. \label{1.15}\eea
The bounds Eq.(\ref{1.15}) are {\it non-perturbative} (i.e., they
are based on the {\it convergence} of the expansion for the
kernels $W^{(h)}$). They are obtained by exploiting the
anticommutativity properties of the Grassmann variables, via a
determinant expansion and the use of the {\it Gram-Hadamard
inequality} for determinants, see \cite{GM}. The factor
$2^{\th h}$ in the bound will play a crucial role in the
following and reflects the fact that the {\it scaling
dimension} $3-n-m$ is always negative for $n>2$. The {\it running coupling constants} $Z_h,v_h,Z_{\m,h}$
satisfy recursive equations ({\it beta function equations}) that,
due to the bound Eq.(\ref{1.15}), lead to bounded and controlled
flows, i.e., $Z(U)=\lim_{h\to-\infty}Z_h$, $Z_\m(U)=\lim_{h\to-\infty}Z_{\m,h}$
and $v_F(U)=\lim_{h\to-\infty}v_h$ are analytic functions of $U$,
analytically close to their unperturbed values $Z_0=Z_{0,0}=1$ and
$Z_{1,0}=Z_{2,0}=v_0=\frac32t$, see \cite{GM}.
The analyticity of
the kernels of the effective potential and of the $h\to-\io$
limits of the running coupling constants implies the analyticity
of the imaginary-time correlation functions (see \cite{GM}) and,
similarly, the analyticity of $\s^{\b,\L}(p_0)$ claimed in the
main theorem.

We are left with proving the universality result Eq.(\ref{1.7}).
To this aim, it is important to notice that $Z_h,v_h,Z_{\m,h}$
are related by Ward Identities. Indeed, proceeding as in
\cite{GMP}, we consider a reference model defined in a way similar
to Eq.(\ref{por}), with the important difference that the
Grassmann integration $P(d\psi)$ is modified into $P_{\ge
h}(d\psi)$, whose propagator differs from the original one by the
presence of a smooth infrared cutoff selecting the momenta $\ge
2^h$; performing the phase transformation $\psi^{\pm}_\xx\to e^{\pm i
\a_\xx}\psi^{\pm}_\xx$ in this functional integral, we find
the analogue of Eq.(\ref{1.10}), which implies
\be \frac{Z_{0,h}}{Z_h}=1+O(U 2^{\th h})\;,\qquad \frac{Z_{1,h}}{Z_h v_h}=
\frac{Z_{2,h}}{Z_h v_h}=
1+O(U 2^{\th h})\label{fon} \ee
where the corrections $O(U 2^{\th h})$ come from the symmetry
breaking terms due to the infrared cut-off function. {\it
Therefore, the effective parameters are
related by exact identities}; the vertex density renormalization
$Z^{(0)}_h$ is equal, up to negligible terms, to the wave function
renormalization, and the current renormalization $Z^{(1)}_h$ is
equal to the product of the effective velocity and the wave
function renormalization \cite{footnote}.

We can write $K_{\m\n}= K_{\m\n}^{(P)}+K^{(D)}_{\m\n}$, where the
two terms in the right hand side correspond to the paramagnetic
and diamagnetic contributions to $K_{\m\n}$, see Eq.(\ref{1.6}).
Note that $\widehat K_{\m\n}^{(D)}(p_0,\vec 0)$ is independent of
$p_0$; using Eq.(\ref{1.15}), we find that $|\widehat
K_{\m\n}^{(D)}(p_0,\vec 0)|\le (\const.) |e|^2 \sum_{h=-\io}^0
2^h$, which is finite. On the other hand $K_{\m\n}^{(P)}(\xx)=$
\bea&& =\sum_{h=-\io}^0 \Big[ 2e^2 \frac{Z_{\m,h} Z_{\n,h}}{(Z_h)^2}
\int\frac{d\kk d\pp}{(2\p)^2|\BBB|^2}e^{i\pp\xx}F_h(\kk,\pp)\cdot\label{1.17}\\
&&\cdot\Tr\{
\G_\m(\vec k,\vec p) C_h(\kk)\G_\n(\vec k+\vec p,-\vec p)C_h(\kk+\pp)\}
+H^{(h)}_{\m\n}(\xx)\Big]\;,\nn\eea
where the first term corresponds to the zero-th order in $U$ in
renormalized perturbation theory ($F_h(\kk,\pp)$ is a suitable smooth cutoff function
constraining $|\kk-\kk_F^\pm|$ and $|\kk-\kk_F^\pm+\pp|$ to be $\simeq 2^h$ and such that $\sum_{h=-\io}^0
F_h(\kk,\pp)=1$; moreover, $Z_hC_h^{-1}(\kk)$ is given by Eq.(\ref{1.8}) with $Z_0,v_0$
replaced by $Z_h,v_h$) and, for all $N\ge 0$ and suitable constants $C_N$,
\be |H^{(h)}_{\m\n}(\xx)|\le C_N |U|\frac{2^{(4+\th) h}}{1+(2^h|\xx|)^N}
\;.\label{1.18} \ee
As compared to the zero-th order contribution to $K^{(P)}_{\m\n}$,
the dimensional bound on $H_{\m\n}^{(h)}$ has an extra factor
$2^{\th h}$, following again from Eq.(\ref{1.15}). From Eq.(\ref{1.18}),
\be |K_{\m\n}^{(P)}(\xx-\yy)|\le(\const.)\frac1{1+
|\xx-\yy|^4}\;,\label{1.19}\ee
that is, $K_{\m\n}(\xx)$ is absolutely integrable and, therefore,
its Fourier transform in the thermodynamic and zero temperature
limit is continuous at $\pp=\V0$. Combining this remark with the
WI Eq.(\ref{1.11}), we find that $\widehat K_{\m\n}(\V0)=0$. In
fact, setting, e.g., $p_2=0$, $\widehat
K_{11}(p_0,p_1,0)=(-p_0/p_1)\widehat K_{01}(p_0, p_1,0)$; taking
first the limit $p_0\to 0$ and then $p_1\to 0$ in the right hand
side, we get $\widehat K_{11}(\V0)=0$; proceeding analogously, we
find that $\widehat K_{\m\n}(\V0)=0$ for all $\m,\n\in\{0,1,2\}$.

On the other hand Eq.\pref{fon} implies that
$\frac{Z_{1,h}}{Z_h}=v_F(U)+ O(U 2^{\th h})$, so that $
K_{\m\n}^{(P)}(\xx)=K_{\m\n}^{(P;0)}(\xx)+K_{\m\n}^{(P;1)}(\xx)$
where $K_{\m\n}^{(P;0)}(\xx)$ is the paramagnetic response
function for the model with Hamiltonian $H^0_\L(\frac23v_F(U))$
and $|K_{\m\n}^{(P;1)}(\xx)|\le
(\const.)|U|(1+|\xx|^{4+\th})^{-1}$, with $0<\th<1$. Therefore, the
Fourier transform of $K_{\m\n}^{(P;1)}$ is {\it differentiable} in
$\pp$ and its derivative is continuous at $\pp=\V0$. A similar
decomposition can be performed in the diamagnetic term, so that,
defining $K^{(1)}_{\m\n}=K^{(P;1)}_{\m\n}+ K^{(D;1)}_{\m\n}$ and
using the WI Eq.(\ref{1.11}), $\sum_{\m=0}^2 p_\m \widehat
K^{(1)}_{\m\n}(\pp)=0$  from which, setting, e.g., $p_2=0$,
we obtain $\widehat K_{11}^{(1)}(p_0,p_1,0)=(p_0/p_1)^2 \widehat
K_{00}^{(1)}(p_0, p_1,0)$; deriving with respect to $p_0$ both
sides and taking first the limit $p_0\to 0$ and next $p_1\to 0$ in
the right hand side, we get $\dpr_{p_0}\widehat
K_{11}^{(1)}(\V0)=0$; proceeding analogously, we find that $
\dpr_{p_\r}\widehat K_{\m\n}^{(1)}(\V0)=0$ for all
$\r,\m,\n\in\{0,1,2\}$. {\it Note the crucial role played by the
continuity of the derivatives of $\hat K^{(1)}_{\m\n}$, which allowed us to exchange
the zero frequency and zero momentum limits}, as compared to the order in Eq.(\ref{1.7}).

In order to compute the conductivity, we are left with the contribution associated
to a free theory with Fermi velocity equal to $v_F(U)$ that, for the 11 component,
setting $v=v_F(U)$, reads:
\bea &&\s_{11}=\lim_{p_0\to 0^+}\lim_{\b\to\infty}\lim_{|\L|\to\infty}\s_{11}^{\b,\L}(p_0)=
2e^2v^2\lim_{p_0\to 0^+}\int\frac{d
k_0}{2\p}\cdot\nn\\
&&\cdot\ \int_{\BBB}\frac{d\vec
k}{|\BBB|}{\rm Tr}\Big\{\frac{S(\kk)-S(\kk+p_0)}{p_0}
\G_1(\vec k,\vec 0)S(\kk)\G_1(\vec k,\vec
0)\Big\}\;.\nn\eea
The latter integral is not uniformly convergent in $p_0$; in
particular, it is well known that one cannot exchange the limit
with the integral \cite{Z}. The integral can be evaluated
explicitly (using residues to compute the integral over $k_0$) and
leads to Eq.(\ref{1.11}). A similar computation shows that
$\s_{22}=\s_{11}$ and that the off-diagonal terms vanish. \qed

The above analysis can be extended to the case of long range
electromagnetic interactions; in such case the wave function,
density and current renormalizations have a strong (anomalous)
power law dependence on the momentum and the Fermi velocity
increases up to the speed of light \cite{GGV,GMP}. WIs similar to
Eq.(\ref{fon}) are still valid and imply that, even if the effective parameters
are strongly momentum dependent, {\it the conductivity only depends weakly on
the frequency} in the optical range.
 
In conclusion, we rigorously proved the non existence of corrections 
to the zero temperature and zero frequency limit of the graphene conductivity
due to weak short range interactions. The proof is based on a combination 
of constructive Renormalization Group methods with exact lattice Ward identities.
Remarkably, this is one of the very few 
examples of universality in condensed matter that can be 
established on firm mathematical grounds. 

\vskip.05truecm A.G. and V.M. gratefully acknowledge financial
support from the ERC Starting Grant CoMBoS-239694. We thank D.
Haldane for valuable discussions on the role of exact lattice Ward
Identities.


\begin{thebibliography}{999999}
\bibitem{footnote}
Eq.(\ref{fon})
should be compared with the result of Theorem 2 in \cite{GM}(b), where
an unnatural definition of the currents was adopted;
this led to an asymmetry
$Z_{1}\not= Z_2$, which is an artifact of the unphysical
definition of the currents and that, correctly, does not show up
here, see \cite{GM}(c).
\end{thebibliography}
\end{document}